\documentclass[
reprint,
superscriptaddress,
aps,
prd,
]{revtex4-2}

\usepackage{graphicx}
\usepackage{dcolumn}
\usepackage{bm}
\usepackage{lineno}
\usepackage{color}
\usepackage{amsmath}
\usepackage{amsfonts}
\usepackage{graphicx}
\usepackage{caption}
\usepackage{lineno}
\usepackage{setspace}
\usepackage{xcolor}
\usepackage[colorlinks,linkcolor=blue,citecolor=blue,urlcolor=blue]{hyperref}
\usepackage[justification=raggedright]{caption}

\begin{document}

\title{The Hadronization Impact on $J/\psi$ Energy Correlators: A Pythia8 Study from Partonic to Hadronic Observables}

\author{Jin-Peng Zhang} 
\affiliation{School of Physics and Information Technology, Shaanxi Normal University, Xi'an 710119, China}  

\author{Qian Yang} 
\email{yangqian2020@sdu.edu.cn (corresponding author)}
\affiliation{Institute of Frontier and Interdisciplinary Science,
Key Laboratory of Particle Physics and Particle Irradiation (MOE),
Shandong University, (QingDao), Shandong 266237, China} 

\author{Wen-Chao Zhang} 
\email{wenchao.zhang@snnu.edu.cn (corresponding author)}
\affiliation{School of Physics and Information Technology, Shaanxi Normal University, Xi'an 710119, China}

\author{Yu-jiao Zhao} 
\affiliation{Institute of Frontier and Interdisciplinary Science,
Key Laboratory of Particle Physics and Particle Irradiation (MOE),
Shandong University, (QingDao), Shandong 266237, China}

\date{\today}

\begin{abstract}
A comprehensive study of the $J/\psi$ energy correlator as a probe of non-perturbative hadronization in color-octet $c\bar{c}$ production is performed.
The energy correlator measures the energy flow as a function of the angular distance ($\chi$) from the identified $J/\psi$ meson.
Using the PYTHIA 8 Monte Carlo event generator, the correlator is computed at both parton and hadron levels. At high $J/\psi$  transverse momentum ($p_T > 7\ \text{GeV}/c$), the parton-level correlator in the $\cos\chi > 0$ region is dominated by soft gluon emission during the hadronization of the color-octet state, a contribution clearly distinguishable from other partonic sources, such as underlying multi-parton interactions. 
The transition to the hadron level, however, introduces substantial modifications, suppressing the correlator in this region by approximately an order of magnitude and underscoring the complexity of the hadronization mapping. 
Further analysis reveals that the hadron-level observable exhibits notable sensitivity to model parameters: increasing the mass splitting between colored $c\bar{c}$ pre-resonances and the $J/\psi$ meson from 0.2 to 0.8 GeV/$c^{2}$ enhances the correlator by up to $60\%$, while extending the color reconnection range yields a milder enhancement of about $10\%$. 
These findings demonstrate that precise measurements of the hadron level $J/\psi$ energy correlator, when interpreted within robust event-generator frameworks, can provide novel constraints on hadronization dynamics and help clarify the production mechanisms of $J/\psi$ state.
\end{abstract}

\maketitle


\section{Introduciton}
The production of $J/\psi$ meson holds a unique and crucial position in testing Quantum Chromodynamics (QCD) \cite{Brambilla:2010cs}.
The large mass of charm quark $(m_{c} \gg \Lambda_{QCD})$ makes the $c\bar{c}$ as a non-relativistic system.
This feature provides a critical theoretical simplification: the heavy mass $m_{c}$ establishes a clear scale hierarchy, cleanly separating the short-distance, perturbative dynamics of the pair's creation from the long-distance, non-perturbative physics of its hadronization. 
This makes $J/\psi$ production an ideal laboratory for probing the interplay between these two fundamental aspects of QCD \cite{Brambilla:2010cs,Braaten:2014ata,KRAMER2001141}.

Several theoretical models form the current landscape for describing $J/\psi$ production, ranging from the foundational color-singlet model \cite{CSM_1975} and the more comprehensive NRQCD factorization approach (which incorporates crucial color-octet states) \cite{NRQCD1,NRQCD2,NRQCD3,NRQCD4}, to the phenomenological color-evaporation model \cite{CEM_2016,CEM_1977} and the high-energy focused \(k_T\)-factorization framework \cite{kT_1,kT_2}. 
Additionally, the parton and hadron cascade modle PACIAE has been applied to study $J/\psi$ production in $proton-proton$ (pp) collisions \cite{PACIAE_1,PACIAE_2}.
Decades of experimental measurements from $e^{+}e^{-}$ colliders \cite{Belle_CX}, fixed-target experiments \cite{NuSea_polar,HERA-B_polar}, and hadron colliders like the Tevatron \cite{CDF_pT,CDF_polar,D0_pT}, RHIC \cite{STAR_pT,Phenix_pT,Phenix_polar,STAR_polar_2014,STAR_polar_2020} and the LHC \cite{LHCb_pT,ALICE_pT,ATLAS_pT1,ATLAS_pT2,ALICE_polar,CMS_Polar} have accumulated a wealth of data on its production cross sections, transverse momentum $(p_{T})$ spectra as well as polarizations.
However, a complete and unified description of $J/\psi$ production remains elusive \cite{NRQCD_Polar1}. 
A persistent challenge lies in simultaneously and accurately describing $J/\psi$ $p_{T}$ spectra and polarization within the same theoretical framework \cite{NRQCD_Polar2, NRQCD3,NRQCD4}.
This tension highlights a fundamental limitation in our current understanding of the final, non-perturbative step: the hadronization of the compact, colored $c\bar{c}$ pair into the observed colorless $J/\psi$ meson.

A promising new approach to directly access this hadronization stage has recently been proposed using the quarkonium energy correlator \cite{JPSIEEC}. This observable, defined in the $J/\psi$'s rest frame (helicity frame), measures the flow of energy deposited at a fixed angular separation from the identified $J/\psi$ momentum direction. 
Its key advantage is infrared safety, ensuring robust predictions for the perturbative QCD (pQCD) component originating from hard gluon emissions during the initial collision. 
Groundbreaking theoretical work has demonstrated that in specific angular regions, particularly near the identified $J/\psi$ angular range, the perturbative radiation is heavily suppressed due to the dead-cone effect and Lorentz boost effect. 
Thus, in this depleted region, the energy correlator becomes sensitive to the non-perturbative soft process. 
Theory predicts that this hadronization contribution can distinctly reveal different color-octect intermediate states, opening a novel window into the transition dynamics \cite{JPSIEEC}.

However, this compelling theoretical prediction immediately highlights a critical gap between theory and experiment. The theoretical calculation provides a parton-level prediction,
while experimental measurements are inevitably conducted at the hadron level, detecting stable particles in the detector such as calorimeter and tracker. The observed final state includes not only partons from the hard process of $c\bar{c}$ production but also those originating from underlying event (UE) activity, such as multiparton interactions (MPI).
Bridging this gap necessitates a realistic modeling of the full event, where the parton-level configuration, including both hard‑scattering products and contributions from MPI, undergoes fragmentation and hadronization into the observed hadronic final state. 
The angular distribution of the predicted hadronization signal in a measurement will be influenced, and potentially significantly altered.
This brings us to the core of the present study: the hadronization impact on the $J/\psi$-energy correlator measurement. 

Pythia 8 serves as a primary event generator for $pp$ collision studies \cite{Pythia81,Pythia82,Pythia83}, renowned for its comprehensive modeling of parton showering, hadronization, and underlying-event dynamics. 
For $J/\psi$ production, Pythia8 implements the NRQCD factorization framework, incorporating both color-singlet and color-octet $c\bar{c}$ production mechanisms. 
This implementation enables the generation of $c\bar{c}$ pre-resonance states. Recent Pythia8-based investigations have successfully examined various $J/\psi$ observables, including polarization effects, multiplicity dependence, and production within jets, providing significant insights into quarkonium production mechanisms \cite{Pythia8_Polar, Pythia_Multi,Pythia_jets}.

For the hadronization of light quarks and gluons, PYTHIA 8 employs the Lund string fragmentation model, wherein confined color fields form strings that fragment via quark–antiquark pair production to yield final-state hadrons. This event generator has achieved remarkable success in describing experimental data from hadron collisions, supported by extensive parameter tuning efforts over the past decades. Various experimental collaborations have performed energy-specific tunings of PYTHIA 8 parameters to match their respective kinematic regimes \cite{Pythia_tune1,Pythia_tune2,Pythia_tune3,Pythia_tune4,Pythia_tune5}; notably, the STAR collaboration recently conducted a dedicated tuning of the underlying-event description for RHIC energies, focusing on adjustments to multi-parton interactions (MPI) and color reconnection (CR) processes \cite{Pythia_tune5}.

The $J/\psi$ energy correlator measurement probes the hadronization of colored $c\bar{c}$ pairs.
However, the hadronization of these pairs is intimately linked to CR and underlying MPI activity via soft gluon emission from color-octet $c\bar{c}$ pre-resonance states. How this coupling shapes the observed correlator remains an open question.
Accurate description of the experimental data requires careful modeling of these effects, with CR being particularly critical.
During hadronization, CR reconfigures color connections among partonic systems to minimize the total string potential energy. This process can alter the spatial and momentum correlations between the $J/\psi$ and the surrounding soft activity—correlations that the energy correlator directly measures.
However, the specific impact of CR on precision quarkonium observables such as the energy correlator remains unexplored. This gap introduces uncertainty when connecting parton-level predictions to hadron-level measurements.

This work presents the first systematic, generator-level study to quantify the impact of hadronization modeling on the newly proposed $J/\psi$ energy correlator. 
The paper is organized as follows. In Section II, we will briefly review the theoretical framework of the $J/\psi$-energy correlator and PYTHIA 8 simulation setup. The main results and their implications for experimental measurements are presented and discussed in Section III.  Finally, Section IV summarizes our findings.

\section{Theoretical Framework and Pythia 8 Simulation Setup}
\subsection{Definition of $J/\psi$-Energy Correlator}

The $J/\psi$-energy correlator is a novel observable proposed as a unique probe of the hadronization dynamics \cite{JPSIEEC}. 
It is designed to measure the energy flow around $J/\psi$ meson in its helicity (rest) frame. 
This measurement provides direct access to the non-perturbative energy deposition associated with the transition of the initial $c\bar{c}$ pair into the final $J/\psi$ hadron.

In the theoretical formulation, the energy correlator is defined at the parton level by summing over the energies of final-state partons. 
To enable a more direct comparison with converntional experimental measurements of energy-energy correlations, in this work we adopt a practical definition that uses the transverse momentum of detected particles as the weighting factor. This approach is applied consistently at both the parton level and the hadron level. 
At the hadron level, the measurement is restricted to charged hadrons. 
This choice is motivated by experimental practicality, as charged-particle tracking offers the most direct and efficient means of measuring both high- and low-energy particles in collider experiments.
Furthermore, since our analysis focuses primarily on soft partonic processes and their transition to hadrons, charged hadrons provide a reliable and well-measured proxy for studying the underlying hadronization dynamics.
The $J/\psi$-energy correlator is thus defined as:
\begin{equation}\label{eq:energy_correlator}
\sum (cos\chi) = \frac{1}{N_{J/\psi}} \sum_{events} \sum_{i \in X} \frac{p_{T,i}}{M_{J/\psi}} \delta (cos\chi - cos\theta_{i}),
\end{equation}
where $\chi$ is the angular distance from the identified $J/\psi$ meson in the helicity frame, $N_{J/\psi}$ is the total number of selected $J/\psi$ events, $p_{T,i}$ is the transverse momentum of the $i$th particle, $M_{J/\psi}$ is the $J/\psi$ mass, and $\theta_{i}$ is the opening angle between the $i$th particle and the $J/\psi$ momentum direction in the $J/\psi$ helicity frame. The summation index $X$ runs over all particles at either parton level or hadron level.

This unified definition preserves the essential sensitivity to hadronization dynamics discussed in theoretical studies, while enabling direct comparisons between parton-level predictions and hadron-level measurements. The region $cos\chi >0$, corresponidng to the forward hemisphere in the $J/\psi$ rest frame, is of particular interest, as perturbative emissions are suppressed due to the dead-cone effect and Lorentz boost effect, potentially enhancing the relative contribution from soft, non-perturbative hadronization processes.

\subsection{Pythia8 Event Generator Configuration}
To simulate the production environment and subsequent hadronization dynamics, we employ the PYTHIA 8.315 event generator. 
Our simulation is designed to provide a baseline that includes the complex interplay of partonic scatterings, shower evolution, and hadron formation, thereby providing a reference against which the predictions of the energy correlator formalism can be tested.
The event generation proceeds as follows. 
Hard scatterings are generated inclusively for $c\bar{c}$ production via the leading-order QCD processes ($gg/q\bar{q}/qg \rightarrow  c\bar{c}+g/q$).
For the production of $J/\psi$ mesons, PYTHIA implements a leading-order (LO) NRQCD framework. 
This framework includes the production of charm quark pairs in both the color-singlet and color-octet configurations, with the latter subsequently evolving into the physical $J/\psi$ state through the emission of a soft gluon. 
The factorized cross section follows the standard NRQCD formalism, summing over relevant intermediate states $n$ (eg. $^{3}S_{1}^{[1]}$, $^{3}S_{1}^{[8]}$,$^{1}S_{0}^{[8]}$,$^{3}P_{J}^{[8]}$) \cite{Pythia83}:
\[\sigma_{pp\rightarrow J/\psi+X} = \sum_{n} \hat{\sigma}_{pp\rightarrow c\bar{c}[n]+X}\langle \mathcal{O}^{J/\psi}[n] \rangle ~. \]
Feed-down contributions from excited charmonium states, such as ${\chi}_{cJ}$, $\psi(2S)$, are also incorporated through their subsequent decays. 
The default PYTHIA parameters for charmonium production, including the relevant long-distrance matrix elements (LDMEs), are employed in this study. 
To ensure realistic modeling of the complete event, the simulation includes the full chain of QCD evolution. This includes initial-state and final-state radiation via the transverse-momentum-ordered parton shower algorithm, UE activity incorporating multiparton interactions and color reconnection effects, as well as the hadronization via the Lund string fragmentation model. All these components are activiated with their repective default settings.

A total of 300 million $J/\psi$ events are simulated in $pp$ collisions at $\sqrt{s}=500$ GeV. 
From this sample, we select events containing at least one $J/\psi$ meson within the kinematic acceptance defined by the transverse momentum $0<p_{T}^{J/\psi}< 15$ GeV/$c$ and rapidity  $|y^{J/\psi}|<1$. 
For the analysis of energy correlator $\sum (cos\chi)$, we consider partons or charged hadrons within the following kinematic range: $p_{T} > 0.2$ GeV/$c$ and pseudorapidity $|\eta|<1$.

\section{RESULTS and DISCUSSION}
\begin{figure}[tbp]
\centering
\includegraphics[scale=0.33]{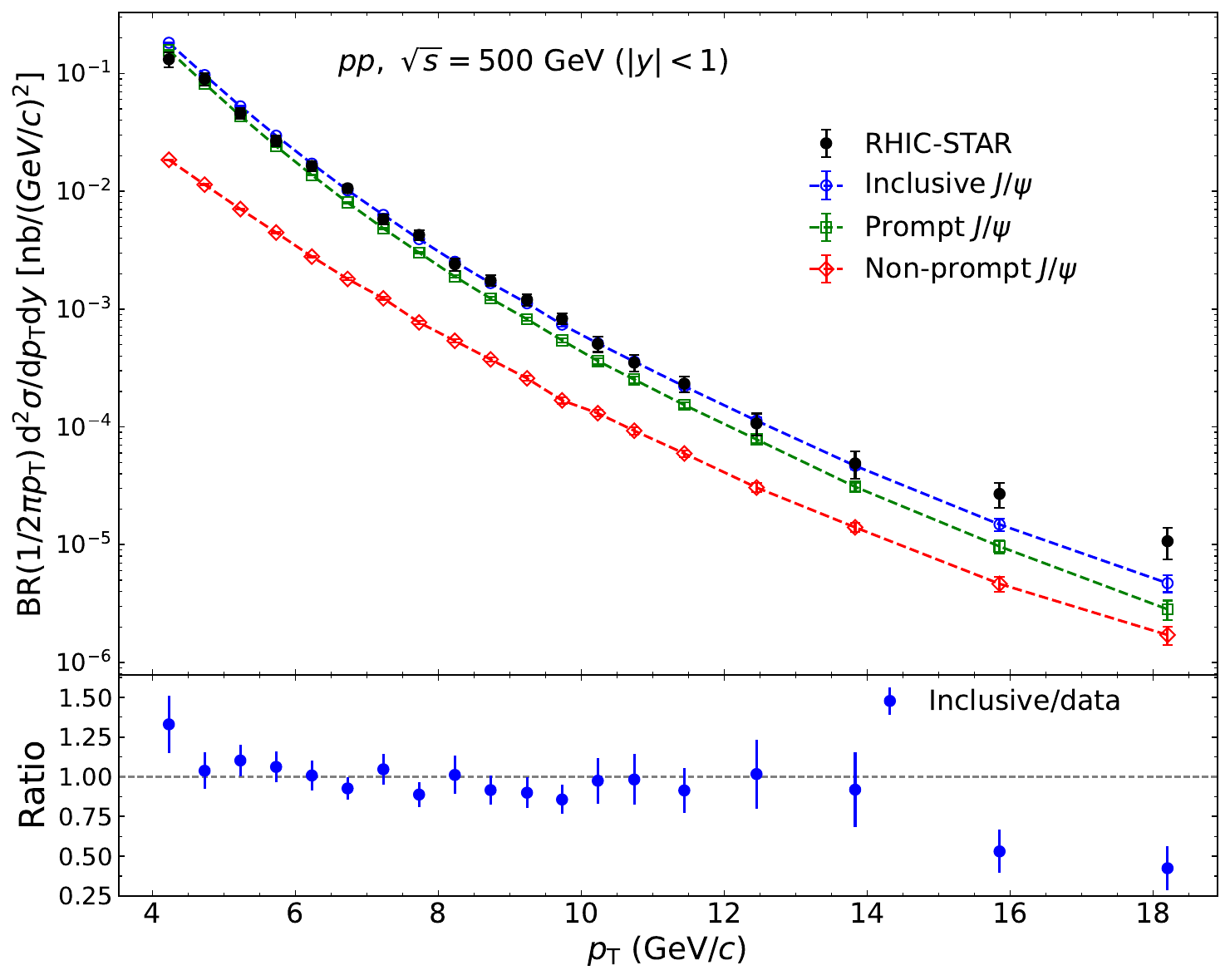}
\caption{Top panel: differential cross sections for $J/\psi$ production in $pp$ collisions at $\sqrt{s}= 500$ GeV \cite{STAR_pT}. The black circles represent the experimental data from RHIC-STAR experiment, while the colored symbols show the PYTHIA results for different $J/\psi$ components: inclusive (blue circles), prompt (green squares), and non-prompt (red diamonds). Bottom panel: ratio of the simulated inclusive $J/\psi$ PYTHIA result to the experiment data.}
\label{fig:CX}
\end{figure}

To access the compatibility of the PYTHIA 8 event generator with experimental data, we first compare the transverse momentum spectrum of the inclusive $J/\psi$ production. In $pp$ collisions at $\sqrt{s}=500$ GeV, the inclusive $J/\psi$ yield predicted by PYTHIA 8 exceeds the corresponding STAR experimental measurement by a factor of 1.88. To enable a meaningful comparison of the spectral shapes, the simulated yields from all $J/\psi$ production processes are uniformly scaled by a factor of $0.53 (1/.1.88)$. 
The upper panel of Fig. \ref{fig:CX} shows the scaled $p_{T}$ spectra from PYTHIA 8 from the inclusive (blue circles), prompt (green squares), and non-prompt (red diamonds) processes from simulation, alongside the RHIC-STAR experimental data (black circles). 
The bottom panel presents the ratio of the scaled inclusive $J/\psi$ spectrum to the data. The comparison reveals that the shape of the inclusive $J/\psi$ $p_{T}$ spectrum from simulation agrees well with the experimental measurement within statistical uncertainties over the studied kinematic range. 
This consistency is very important for the subsequent analysis of $J/\psi$-energy correlators, as the measurement is performed in the $J/\psi$ rest frame, requiring a Lorentz boost of all the associated partons and hadrons.

\begin{figure*}[htbp]
\includegraphics[scale=0.395]{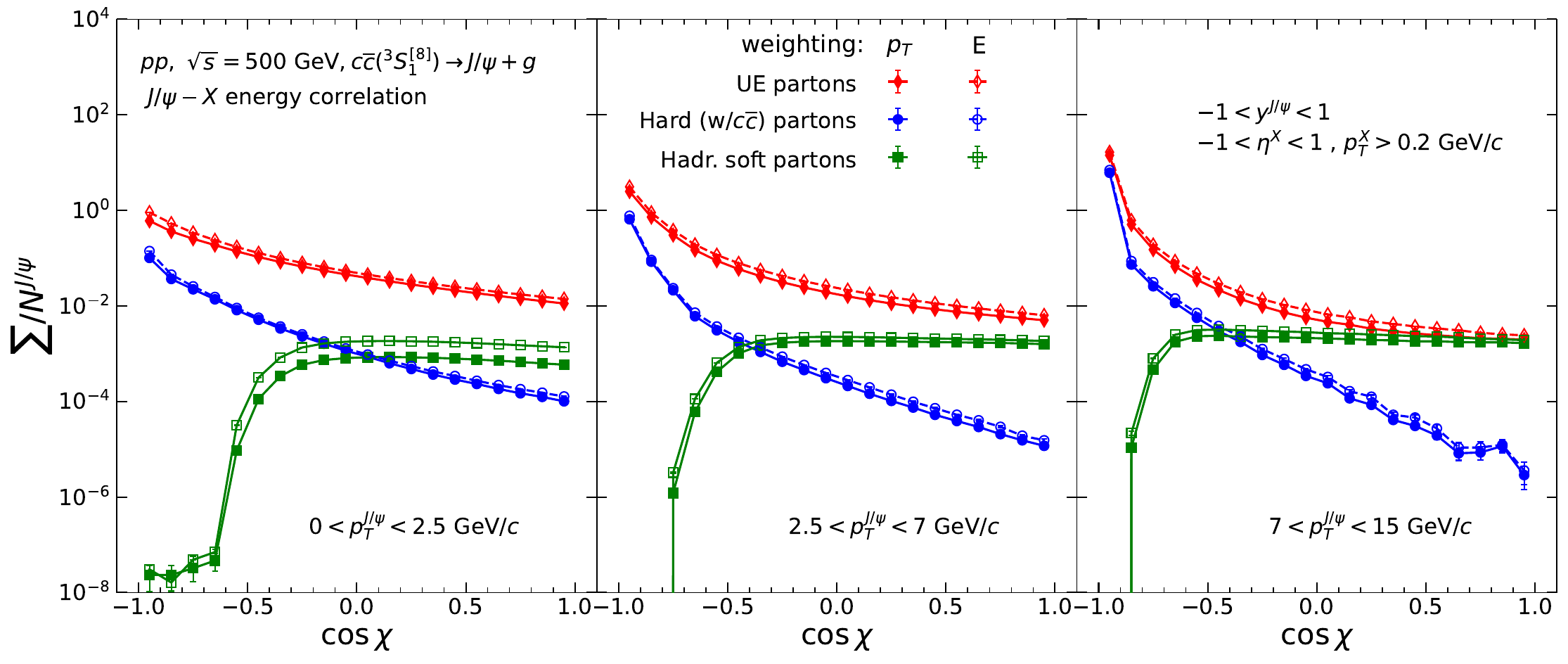}
\caption{\label{fig:wide} $J/\psi$-energy correlator at parton level for events with $^{3}S_{1}^{[8]}\rightarrow J/\psi + gluon$ process, including contribution from: Underlying-event (UE, red diamonds), hard-scattered partons produced in association with the $c\bar{c}$ pair via $q\bar{q}/gg/qg \rightarrow c \bar{c}^{[8]} + g/q$ (Hard (w/c $c\bar{c}$),blue circles), and soft partons emitted during the $c\bar{c}$ hadronization process (Hadr. soft, green squares). Solid markers correspond to results using transverse-momentum weighting, while open markers represent results using parton energy weighting for the correlator definition. Results are shown in three $J/\psi$ $p_{T}$ intervals: $0 < p_{\mathrm{T}}^{J/\psi} < 2.5$~GeV/$c$ (left), $2.5 < p_{\mathrm{T}}^{J/\psi} < 7$~GeV/$c$ (middle), and $7 < p_{\mathrm{T}}^{J/\psi} < 15$~GeV/$c$ (right). Kinematic requirements of $-1 < y^{J/\psi} < 1$, $-1 < \eta^X < 1$, and $p_T^X > 0.2$ GeV/$c$ are applied.}
\label{fig:parton_level_eec}
\end{figure*}

To investigate the behavior of the $J/\psi$-energy correlator at the parton level, we focus on the energy flow associated with the hadronization of color-octet $c\bar{c}$ pairs into $J/\psi$. 
In PYTHIA 8, this non-perturbative process is primarily modeled via intermediate color-octet $c\bar{c}$ states, (mainly $^{3}S_{1}^{[8]}$, $^{1}S_{0}^{[8]}$, and $^{3}P_{J}^{[8]}$), which share a common mass defined as the $J/\psi$ mass plus the parameter `Onia:massSplit'. 
These states subsequently hadronize into a $J/\psi$ by radiating a soft gluon. 
In this work, partons are categorized into three classes: 1) the soft gluon emitted during the $c\bar{c}$ hadronization (`Hadr. soft'); 2) partons produced in association with the $c\bar{c}$ pair from the primary hard scattering (`Hard (w/ $c\bar{c}$)'); 3) partons from the UE, predominantly originating from MPI processes.
Assuming that underlying MPI contributions are largely independent of the specific hard processes that produce the $c\bar{c}$, we use events generated via the $c\bar{c} (^{3}S_{1}^{[8]})$ channel as a representative case to illustrate the partonic energy flow.

Fig. \ref{fig:parton_level_eec} presents the parton-level $J/\psi$-energy correlator as a function of $\cos\chi$ for events including the $c\bar{c}(^{3}S_{1}^{[8]}) \rightarrow J/\psi + g$ process. It is decomposed into three categories across three $J/\psi$ $p_T$ intervals. 
In order to be consistent with the typical experimental acceptance, the following kinematic requirements $|y^{J/\psi}|<1$, $|\eta^X|<1$, and $p_T^X > 0.2$ GeV/$c$ are applied. 
The distribution arising from  the hadronization soft gluons (green squares) displays a distinct kinematic structure: a sharp cutoff followed by a plateau. 
This patten originates from the isotropic emission of a fixed-energy gluon in the rest frame of the $c\bar{c}$ state, which is then modified by the Lorentz boost to the $J/\psi$ rest frame and further shaped by the applied parton $p_T$ and $\eta$ cuts.
In contrast, the correlator shapes for associated hard-scattering partons (blue circles) and UE partons (red diamonds) show a pronounced shift toward $\cos\chi < 0$ as the $J/\psi$ $p_T$ increases.

Crucially, for $J/\psi$ $p_T > 7$ GeV/$c$, the contribution from the soft gluon emitted during hadronization remains significant in the region with $\cos\chi > 0$, even after including the contributions from UE partons and associated hard partons. In this region, approximately $39.3\%$ of the energy flow stems from the soft hadronization process. This fraction increases to about $44.5\%$ for $\cos\chi > 0.5$. This finding demonstrates that contribution from UE MPI processs is also suppressed in the $\cos\chi > 0$ region due to the boost effect, thereby providing a valuable kinematic window for isolating and studying the energy carried by the soft gluon released during $J/\psi$ formation.

In Eq. (\ref{eq:energy_correlator}), the transverse momentum of particles is adopted as the weighting factor instead of the particle energy, leading to a minor deviation of the present definition from the theoretical formulation proposed in Ref. \cite{JPSIEEC}. Given that comparisons with theoretical predictions are equally crucial for this new observable, an additionally investigated the parton-level $J/\psi$ energy correlator using the parton energy as the weighting factor is carried out, with the corresponding results presented by the open markers in Fig. \ref{fig:parton_level_eec}. 
The correlator shows a slight enhancement when parton energy is used as the weight, and both definitions exhibit qualitatively identical kinematic trends across all studied regions. In addition, for the soft gluon distribution from $c\bar{c}$ hadronizationthe, the quantitative difference between the $p_T$-weighted and energy-weighted results shows a dependence on the $J/\psi$ transverse momentum. This momentum dependence can be attributed to the decay dynamics of the $c\bar{c}$ system and the associated Lorentz boost effect. 

\begin{figure*}[htbp]
\centering 
\includegraphics[scale=0.395]{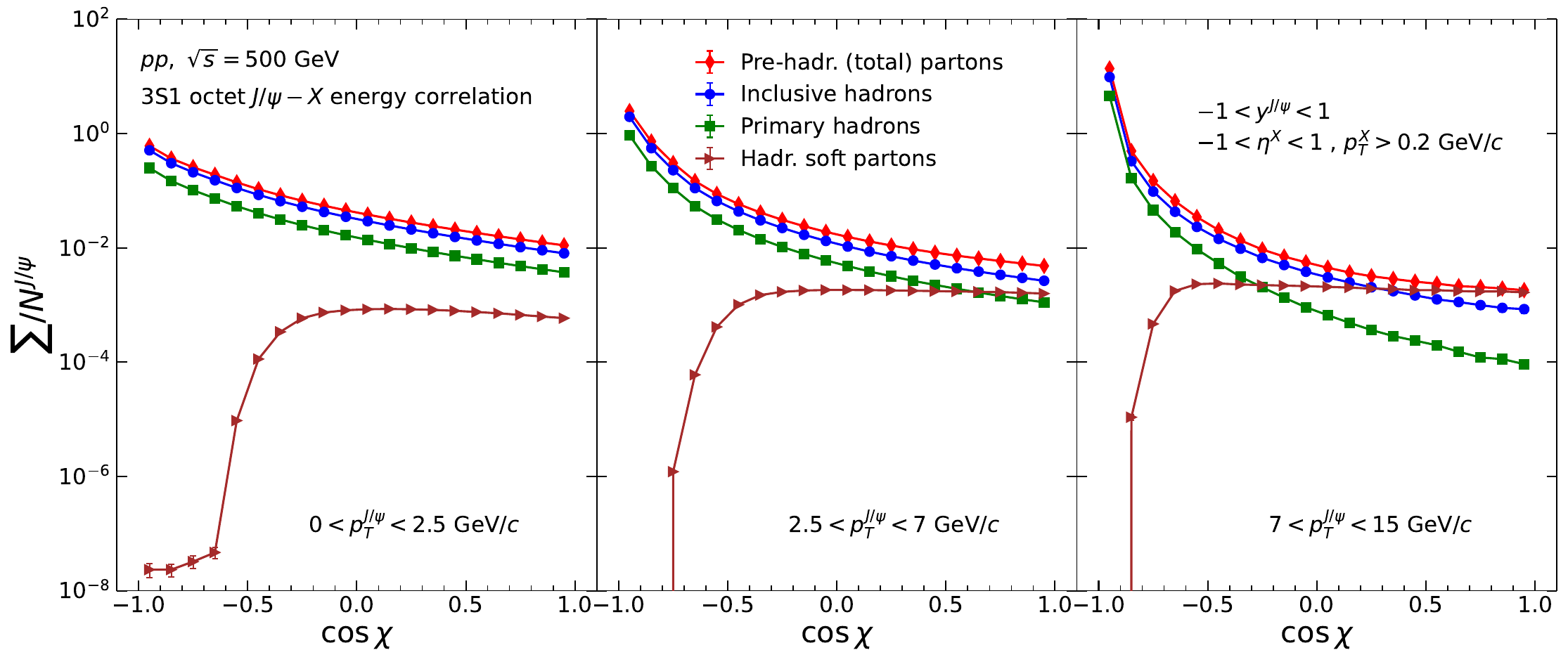}
\caption{\label{fig:wide} $J/\psi$-energy correlator at the hadron and parton level for events with $c\bar{c}(^{3}S_{1}^{[8]})\rightarrow J/\psi + gluon$ process. Contributions are shown from all pre-hadonization partons (Pre-hadr.(total) partons), primary hadrons (Primay hadron, red circle), inclusive hadrons that additionally include hadron decay contributions (inclusive hadron, blue triangles), and soft partons released during the $c\bar{c}$ hadronization process (Hadr. soft,green squares). Results are presented in three $J/\psi$ $p_{T}$ intervals: $0 < p_{\mathrm{T}}^{J/\psi} < 2.5$~GeV/$c$ (left), $2.5 < p_{\mathrm{T}}^{J/\psi} < 7$~GeV/$c$ (middle), and $7 < p_{\mathrm{T}}^{J/\psi} < 15$~GeV/$c$ (right). Kinematic requirements of $-1 < y^{J/\psi} < 1$, $-1 < \eta^X < 1$, and $p_T^X > 0.2$ GeV/$c$ are applied.}
\label{fig:hadron_vs_parton}
\end{figure*}

The hadronization of the soft gluon, as well as other partons, involves complex final-state interactions to balance energy and color charge. 
In this study, PYTHIA 8 models this process by the Lund string fragmentation framework, with MPI-based color reconnection enabled. 
Fig. \ref{fig:hadron_vs_parton} presents the $J/\psi$-energy correlator at the both parton and hadron levels for directly produced $J/\psi$ in the same three $p_T$ intervals as shown in Fig. \ref{fig:parton_level_eec}. 
The correlators include: all partons before hadronization (`pre-hadr. (total) partons', red diamonds), primary hadrons (`Primary hadron', green squares), inclusive hadrons which include decay products (`Inclusive hadron', blue circles), and the soft partons specifically from the $c\bar{c}$ hadronization (`Hadr. soft', magenta triangles) as a reference.

Comparing the energy correlator of partons before hadronization with that of primary hadrons reveals several interesting features. 
First, the characteristic enhancement in the $\cos\chi < 0$ region, induced by the boost to the $J/\psi$ rest frame, persists at the hadron level, indicating that the hadronization process does not erase this kinematic signature. 
Second, the primary hadron correlator is systematically lower than its partonic counterpart across the entire $\cos\chi$ range, which is expected as a portion of the partonic energy is converted into hadron rest masses. 

At high $J/\psi$ $p_{T}$ region, the difference between the parton-level and hadron-level energy flow becomes increasingly pronounced in the region of interest, $\cos\chi > 0$. 
Near $\cos\chi \sim 1$, the hadronic correlator is suppressed by more than an order of magnitude relative to the partonic energy correlator, demonstrating that the hadronization process drastically reshapes the energy flow distribution, and strongly suppresses the hadron production in this kinematic region, despite significant partonic energy flow presents.

The inclusive hadron correlator (including decay products) shows an overall increase compared to the primary hadron correlator. 
This increase is attributed to the decay process converting rest mass into kinetic energy, thereby adding additional particles to the energy flow. 
The enhancement appears largely uniform across $\cos\chi$, indicating that decays provide a roughly constant additive contribution to the energy correlator over the measured angular range.

In the $J/\psi$ prodcution models, the precise amount of energy released during the non-perturbative transition of a color-octet $c\bar{c}$ pair to a $J/\psi$ meson remains an open question. In NRQCD approach, this energy is predominantly carried away by soft gluon radiation, and represents a crucial aspect of the hadronization dynamics. 
To explore how such energy release appears in an experimentally accessible observable, we study the sensitivity of the hadron-level $J/\psi$ energy correlator to variations in this parameter within the simulation framework. In PYTHIA 8, the mass difference between the initial color-octet $c\bar{c}$ state and the final $J/\psi$ is controlled by the parameter `Onia:massSplit', which effectively sets the energy scale available for the soft gluon emission during the transition.

\begin{figure*}[htbp]
\centering
\begin{minipage}{0.47\textwidth}
\centering
\includegraphics[scale=0.33]{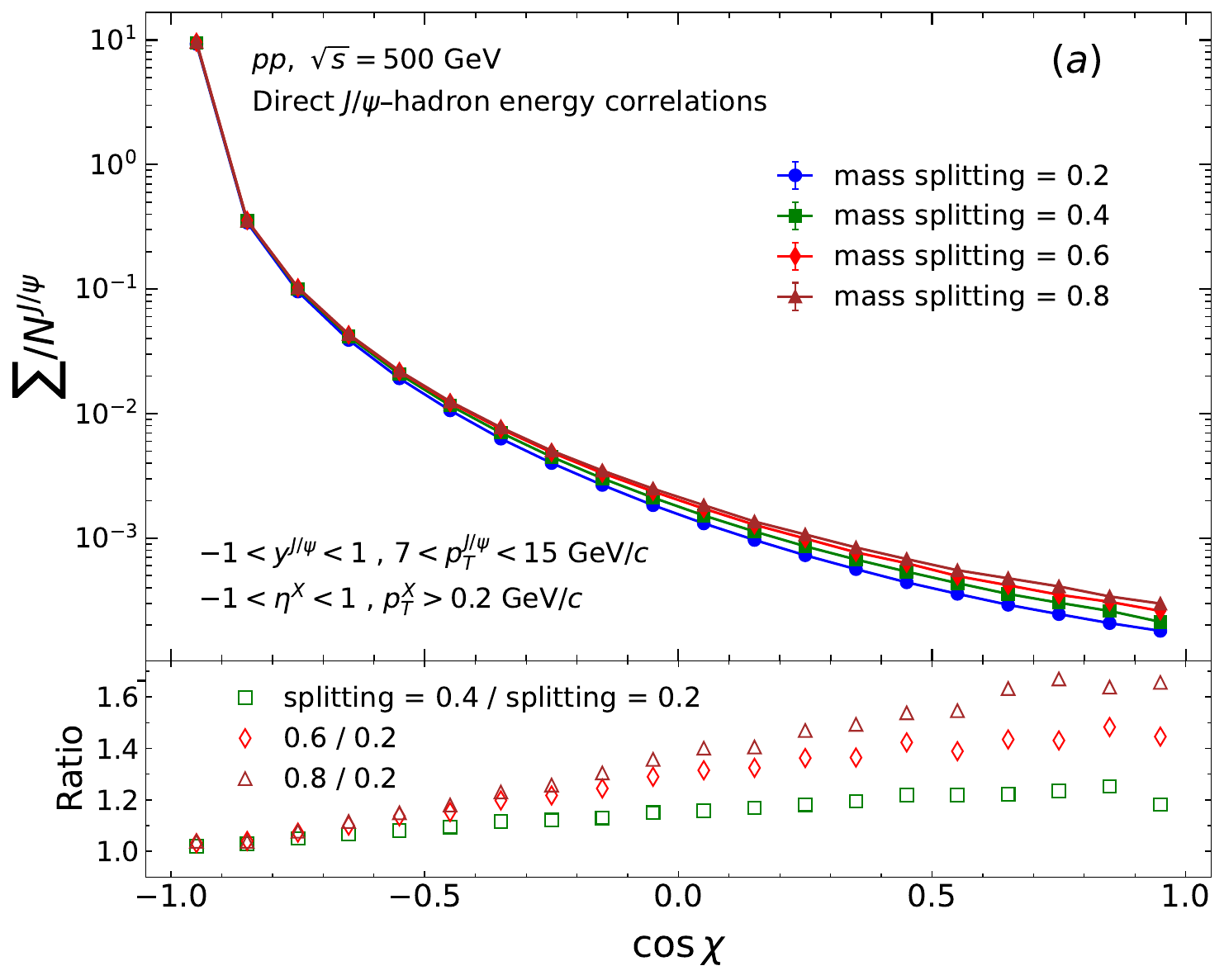}
\end{minipage}
 \hspace{0.03\textwidth}
\begin{minipage}{0.47\textwidth}
\centering
\includegraphics[scale=0.33]{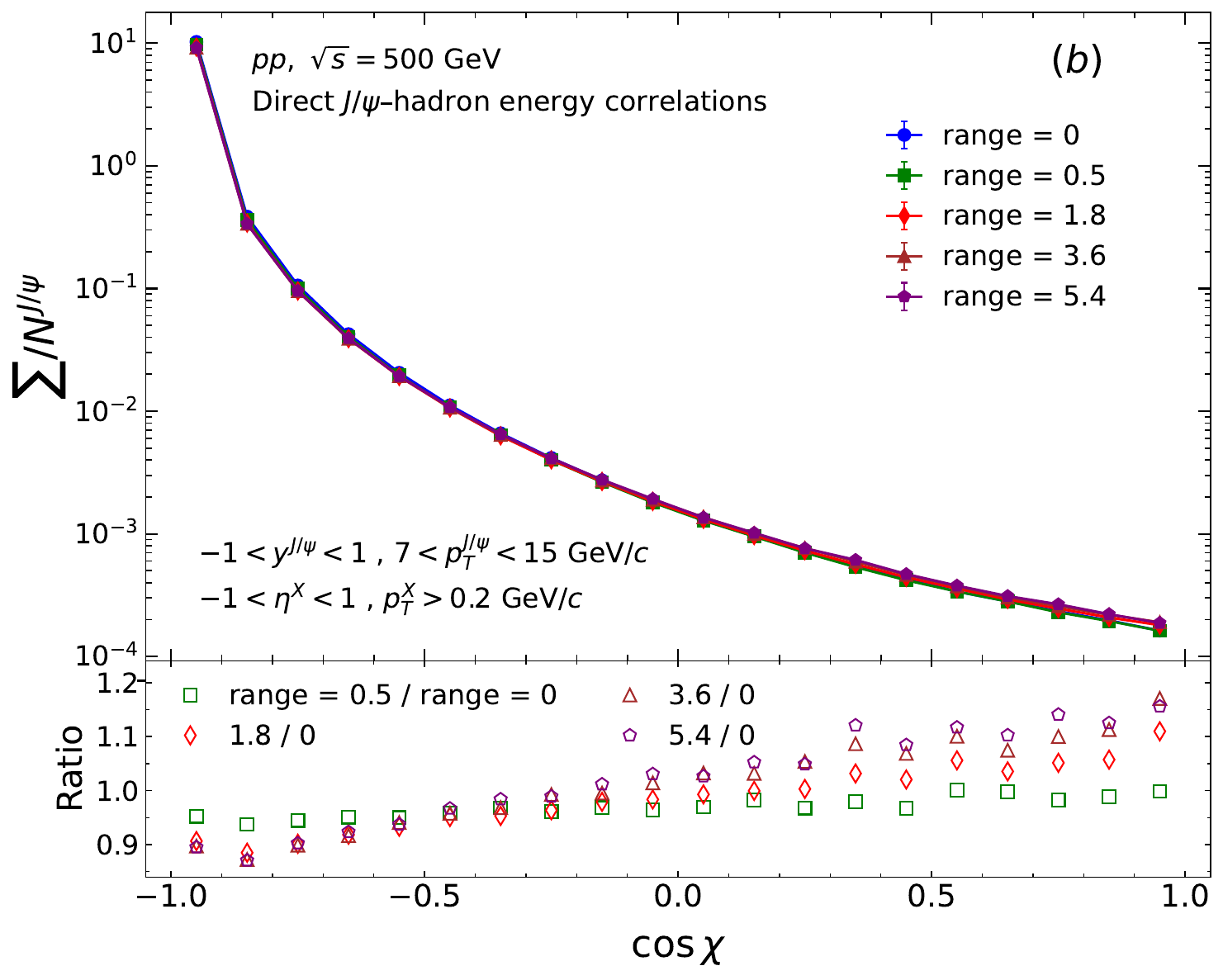}
\end{minipage}

\caption{
The hadron level $J/\psi$ energy correlator as a function of $cos \chi$ for the directly produced $J/\psi$ production in pp collisions at $\sqrt{s}=500$ GeV. The $J/\psi$ is selected with $|y^{J/\psi}|<1$ and $7 < p_{T}^{J/\psi} < 15$ GeV/$c$. Associated charged particles $X$ satisfy  $-1 < \eta^X < 1$, and $p_T^X > 0.2$ GeV/$c$.
(a) The top panel shows the dependence on the color-octet $c\bar{c}$ mass splitting parameter 0.2 (blue circles), 0.4 (green
squares), 0.6 (red diamonds), and 0.8 (magenta triangles) GeV/$c^{2}$. The bottom panel shows the ratio of each distribution to the default case of 0.2 GeV/$c^{2}$.
(b) The top panel shows the dependence on the color reconnection range parameter R = 0 (blue circles), 0.5 (green squares), 1.8 (red diamonds), 3.6 (magenta triangles), and 5.4(purple pentagon), with Onia:massSplit fixed at 0.2 GeV/$c^{2}$. 
The bottom panel shows the ratio of each distibution to the $R=0$ case. 
}
\label{fig:MS_CR}

\end{figure*}

The top panel of Fig. \ref{fig:MS_CR}(a) presents the hadron-level $J/\psi$ energy correlator for directly produced $J/\psi$ in $pp$ collisions at $\sqrt{s}=500$ GeV, with $J/\psi$ selected using $|y^{J/\psi}|<1$, $7 < p_T^{J/\psi} < 15$ GeV/$c$, and associated charged particles satisfying $-1 < \eta < 1$, $p_T > 0.2$ GeV/$c$. The energy correlator is shown for four values of `Onia:massSplit': 0.2, 0.4, 0.6, and 0.8 GeV/$c^2$. The top panel reveals a clear systematic trend: as the mass splitting (and thus the released energy) increases, the correlator in the region with $\cos\chi > 0$ is enhanced. 
The bottom panel quantifies this enhancement by showing the ratio of each distribution to the default case of 0.2 GeV/$c^2$. The effect is substantial. The largest mass splitting of 0.8 GeV/$c^2$ leads to an increase of up to approximately $60\%$ in the $\cos\chi > 0$ region compared to the default.
This suggests that the additional energy from the hadronization process materializes as an extra hadronic activity preferentially aligned with the $J/\psi$ flight direction in the boost frame.

To further probe the non-perturbative aspects of the $c\bar{c}$ hadronization, we investigate the influence of Colour Reconnection on the hadron-level $J/\psi$ energy correlator. CR is a key element of the Lund string fragmentation model, describing the rearrangement of colour charges among final-state partons to form colour-singlet strings prior to hadronization. In the specific process where a colour-octet $c\bar{c}$ pair emits a soft gluon to become a colour-singlet $J/\psi$, the emitted gluon carries both energy and colour. Through CR, this gluon can be incorporated into a string system that may also include soft partons from UE or MPI. The spatial range over such reconnections are allowed, controlled by the ‘ColourReconnection:range’ parameter ($R$) in PYTHIA 8, could thus modify the final hadronic energy flow distribution around the $J/\psi$.

The top panel of Fig. \ref{fig:MS_CR}(b) presents the influence of the different CR range scenarios on $J/\psi$ energy correlators, with the `Onia:massSplit' parameter is fixed at 0.2 GeV/$c^2$ to isolate a specific hadronization energy scale.
The red diamonds, magenta triangles, blue circles, and green squares, respectively,  represent the cases with $R$=0, 0.5, 1.8 and 3.6.
The lower panel quantifies these effects by presenting the ratio of each correlator to the $R=0$ baseline.
In the $\cos\chi < 0$ region, the ratio is consistently less than unity for all $R>0$: $R=0.5$ results in a relatively higher ratio, while $R=$1.8, 3.6, and 5.4 exhibit overlapping distribution shapes with ratios lower than that of $R=0.5$. While in the $\cos\chi > 0$ region, the $R=0.5$ results align nearly identically with the $R=0$ baseline. A clear enhancement in the ratio only arises when $R$ exceeds 1.8, and this enhancement becomes progressively more pronounced as $R$ increases (from 1.8 to 5.4). Notably, all quantitative differences between CR-enabled scenarios and the $R=0$ baseline are constrained within $\sim 10\%$.

This pattern indicates that CR modifies the string topology, effectively redistributing the energy flow from fragmenting systems and altering the angular correlation between the $J/\psi$ and the bulk of hadronic energy. 
The $R$-dependence in the $\cos\chi > 0$ region may be attributed to the lower parton density in this kinematic domain. Here, soft gluons need a sufficiently large CR range to connect with distant color sources, forming longer strings that undergo hadronization, thereby inducing the observed $R$-dependent enhancement.

A key finding from Fig. \ref{fig:MS_CR}(b) is the interplay between CR range $R$ and the correlator: while qualitative trends (suppression in $\cos\chi < 0$ for all $R>0$) are robust to $R$, quantitative behaviors (enhancement in $\cos\chi > 0$) are triggered above the default $R$ (1.8) and strengthen with increasing $R$. Nevertheless, the overall variation remains within  $\sim 10\%$. This implies that the core energy flow modification from CR, namely the momentum smearing via embedding $c\bar{c}$ radiated soft gluon into a broader color network, is a robust outcome of mechanism activation with mild quantitative dependence on the precise spatial scale. Such stability suggests limited theoretical uncertainties from $R$ modeling for this observable.

Importantly, despite the significant morphological differences between parton and hadron level distributions, the hadron level correlator retains and reflects the characteristic energy scale of the non-perturbative hadronization process. 
This is further evidenced by the enhancement in the $\cos\chi>0$ region when either the mass splitting between the colored $c\bar{c}$ states is increased or the CR range $R$ is enlarged. The observed sensitivity provides a novel and phenomenologically valuable handle to constrain poorly determined hadronization parameters. 
Consequently, precise experimental measurements of this correlator could help disentangle the contributions from different $J/\psi$ production mechanisms and refine our understanding of the transition dynamics from colored heavy-quark states to color-singlet hadrons. Future comparisons with data will therefore be essential to validate and tune these parameters, thereby reducing theoretical uncertainties in $J/\Psi$ production phenomenology.

The present analysis highlights a key challenge for experimental measurements of the $J/\psi$ energy correlator: while the parton-level dynamics are of primary theoretical interest, only the hadron-level energy correlator is directly accessible in collider experiments. 
As illustrated in Fig. \ref{fig:hadron_vs_parton}, significant reshaping of the energy-flow distribution by the hadronization process, makes a straightforward inversion from the hadronic to the partonic observable unfeasible.
A pragmatic and effective approach is therefore to employ event generators, such as PYTHIA 8, to establish a calibrated mapping between the measurable hadron level correlator and the parton level dynamics of interest. 
By comparing experimental data with generator predictions at the hadron level, which inherently include modeling of fragmentation, hadronization, and decays, one can constrain the generator parameters and the underlying physics models. 
The validated generator can then be used to extract parton level information, such as the energy flow of the soft hadronization gluon, from the measured distributions.

Our study based on PYTHIA 8 clearly demonstrates that non-perturbative hadronization dynamics play a crucial role in shaping the $J/\psi$ energy correlator, especially the significant suppression of the correlator in the $\cos\chi>0$ region at the hadron level relative to the parton level. It is worth noting that the quantitative description of hadronization effects is inherently dependent on the specific modeling of quarkonium production and fragmentation processes in event generators, and thus complementary studies using different event generator frameworks are highly beneficial for a comprehensive understanding of the hadronization impact on the $J/\psi$ energy correlator. The latest version of HERWIG 7 \cite{herwig} features sophisticated modeling of quarkonium production processes, including the dedicated quarkonium parton shower implementation following the NRQCD formalism, which provides an important and beneficial framework for further in-depth studies of the $J/\psi$ energy correlator. A systematic investigation of the $J/\psi$ energy correlator using HERWIG 7 will be carried out in our future work to obtain a more comprehensive understanding of the hadronization effects on this observable.

Our discussion has so far focused on the energy correlator for directly produced $J/\psi$ mesons. For $J/\psi$ mesons originating from the decay of excited charmonium states, the hadron-level energy correlator will receive additional modifications from the decay products. 
In PYTHIA 8, the color-octet $c\bar{c}$ hadronization to states like $\psi(2S)$ and $\chi_{cJ}$ is modeled similar to that of $J/\psi$. 
However, their subsequent decays imprint distinct features: $\psi(2S) \rightarrow J/\psi \, \pi^+ \pi^-$ decays enhance the correlator in the $\cos\chi > 0$ region compared to direct production, due to the added charged pions. 
In contrast, $\chi_{cJ} \rightarrow J/\psi \, \gamma$ decays, where the photon is not included in the energy correlator measured with charged hadrons only, yield a distribution much closer to that of direct the $J/\psi$ production. 
Finally, $J/\psi$ mesons from $b$-hadron decays are expected to show a significant enhancement at $\cos\chi > 0$, primarily due to the fragmentation products of the $b$ quark.

\section{Summary}
This work presents the first systematic, generator-level study of the $J/\psi$ energy correlator, quantifying its sensitivity to the non-perturbative hadronization dynamics of color-octet $c\bar{c}$ states. The analysis is performed using PYTHIA 8 event generator, which implements the NRQCD framework. At the parton level, the soft gluon released during the $c\bar{c}$ hadronization process constitutes a significant component of the energy flow in the $\cos\chi > 0$ region at high $J/\psi$ $p_{T}$. This contribution remains substantial and can be distinguished from the underlying event and hard process. 
A critical observation from the comparison of parton and hadron level distributions is that the correlator is dramatically reshaped during hadronization.
Its magnitude in the theoretically pivotal $\cos\chi > 0$ region is suppressed by approximately an order of magnitude at the hadron level. 
This pronounced difference highlights the necessity of a reliable hadron-to-parton mapping for extracting the gluon's energy from experimental data. 
Furthermore, the impact of the key hadronization model parameters on the $J/\psi$ energy correlator is investigated. 
Increasing the mass splitting between the colored $c\bar{c}$ pre-resonance states and $J/\psi$ enhances the hadron-level correlator at $\cos\chi > 0$ by up to $\sim 60\%$, while enlarging the color-reconnection range leads to a more modest enhancement of $10\%$. 
In summary, this analysis establishes the hadron-level $J/\psi$-energy correlator as a sensitive and viable probe for empirical studies of the soft gluon emission and color neutralization in $J/\psi$ formation.

\begin{acknowledgments}
This work is supported in part by the National Key Research and Development Program of China under Contract No.2022YFA1604900 and by the National Natural Science Foundation of China under Grant Nos. 11447024, 11505108 and 12105155.
\end{acknowledgments}



\bibliography{apssamp}

\end{document}